\documentclass[english]{revtex4}
\usepackage[T1]{fontenc}
\usepackage{mathrsfs}
\usepackage{amsmath}
\usepackage{amssymb}
\usepackage{stmaryrd}

\makeatletter
\@ifundefined{textcolor}{}
{%
 \definecolor{BLACK}{gray}{0}
 \definecolor{WHITE}{gray}{1}
 \definecolor{RED}{rgb}{1,0,0}
 \definecolor{GREEN}{rgb}{0,1,0}
 \definecolor{BLUE}{rgb}{0,0,1}
 \definecolor{CYAN}{cmyk}{1,0,0,0}
 \definecolor{MAGENTA}{cmyk}{0,1,0,0}
 \definecolor{YELLOW}{cmyk}{0,0,1,0}
}

\makeatother

\usepackage{babel}
\begin{document}
\title{Well-posedness of Ricci Flow in Lorentzian Spacetime and its Entropy
Formula}
\author{M.J.Luo}
\address{Department of Physics, Jiangsu University, Zhenjiang 212013, People's
Republic of China}
\email{mjluo@ujs.edu.cn}

\begin{abstract}
This paper attempts to construct monotonic entropy functionals for
four-dimensional Lorentzian spacetime under physical boundary conditions,
as an extension of Perelman's monotonic entropy functionals constructed
for three-dimensional compact Riemannian manifolds. The monotonicity
of these entropy functionals is utilized to prove the well-posedness
of applying Ricci flow to four-dimensional Lorentzian spacetime for
a long flow-time, particularly for the timelike modes which would
seem blow up and ill-defined. The general idea is that the Ricci flow
of a Lorentzian spacetime metric and the coupled conjugate heat flow
of a density on the Lorentzian spacetime as a whole turns out to be
the gradient flows of the monotonic functionals for a long flow-time,
so the superficial ``blow-up'' in the individual Ricci flow system
or the conjugate heat flow system contradicts the boundedness of the
monotonic functionals within finite flow interval, which gives a semi-global
control to the whole coupled system. The physical significance and
applications of these monotonic entropy functionals in real gravitational
systems are also discussed.
\end{abstract}
\maketitle

\section{Introduction}

The Ricci flow is a geometric flow proposed by Hamilton \citep{Hamilton1982Three,hamilton1986four},
which is a continuous deformation of a Riemannian geometric metric
$g_{ij}$ driven by the Ricci curvature $R_{ij}$
\begin{equation}
\frac{\partial g_{ij}}{\partial t}=-2R_{ij}\label{eq:3d Ricci flow}
\end{equation}
This equation was also independently discovered in physics by Friedan
\citep{friedan1980nonlinear,Friedan1980}. This equation is a weakly
parabolic type equation. Later, DeTurck proposed applying an appropriate
diffeomorphism to the right-hand side of the equation, which could
transform it into a strongly parabolic one. This approach is now known
as the DeTurck trick \citep{deturck1983deforming}, and it facilitated
the proof of the existence and uniqueness of solutions for this type
of equation (the Ricci flow equation after DeTurck's deformation).

This equation plays a pivotal role in proving the Poincar\'{e} Conjecture
and Thurston's Conjecture in three-dimensional compact Riemannian
geometry. The Ricci flow can gradually deform any initial three-dimensional
Riemannian manifold, and when singularities emerge during this deformation
process, a series of theorems introduced by Perelman \citep{perelman2002entropy,perelman2003ricci,perelman307245finite}
can be employed to remove local singularities through surgeries. Subsequently,
the continuous deformation process via the Ricci flow can restart,
enabling any initially given Riemannian manifold to gradually evolve
into one of the eight possible fundamental three-dimensional Riemannian
manifolds conjectured by Thurston. As a special case within this framework,
the Poincar\'{e} Conjecture has also been proven.

Although the Ricci flow has achieved tremendous success in three-dimensional
Riemannian geometry, it is generally considered ill-defined when applied
to four-dimensional Lorentzian manifolds (the geometry of real spacetime).
The reason lies in the fact that, under the Lorentzian signature $(-,+,+,+)$,
unlike the spacelike modes of metric where high-frequency modes are
gradually attenuated by the parabolic equation, the timelike modes
metric renders the Ricci flow equation no longer a simple parabolic
equation, so that the high-frequency modes of such an equation will
grow exponentially, rendering the solution unstable, often called
``high-frequency blow-up''. Therefore, the Ricci flow for the metric
of timelike modes are backward parabolic, posing severe well-posedness
issues. 

Recently, within a quantum spacetime reference frame theory (a type
of nonlinear sigma model) that we have proposed \citep{Luo2014The,Luo2015Dark,Luo:2015pca,Luo:2019iby,Luo:2021zpi,Luo:2022goc,Luo:2022statistics,Luo:2022ywl,Luo:2023eqf,2023AnPhy.45869452L,2024arXiv240809630L},
the renormalization of the quantum reference frame can be regarded
as the renormalization group (RG) flow of the frame fields (scalar
fields) on a laboratory flat spacetime (the base spacetime). Under
the quantum equivalence principle, the Ricci flow can also be equivalently
viewed as a RG flow of a Lorentzian curved spacetime (the target spacetime)
measured by the configuration of the frame fields, at the one-loop
level. Thus, the Ricci flow of a Lorentzian curved spacetime has a
physical foundation: it provides a process that, under the relativistic
premise of preserving the local speed of light and causal structure,
gradually averages out the fine-grained structure of spacetime through
quantum corrections and renormalization at small scales, transforming
it into a coarse-grained structure.

Many studies \citep{1984Smoothing,Piotrkowska:1995yx,Buchert:2002ht,Carfora:2008fr}
in the literature also explore applying the Ricci flow solely to spacelike
hypersurfaces within four-dimensional spacetime (where time does not
follow the Ricci flow) or to four-dimensional Euclidean spacetime
\citep{2006Ricci} (where Euclidean time allows for the normal application
of the conventional Ricci flow), in order to circumvent the ill-posedness
issues associated with applying the Ricci flow to Lorentzian spacetime.
However, these attempts fail to preserve the causal structure of real
spacetime and can thus only be regarded as approximations. Currently,
there are only a few specific examples of the Ricci flow for maximally
symmetric Lorentzian spacetimes in the literature \citep{2020Constructing,Cartas-Fuentevilla:2017cvt},
which at least demonstrates that, in certain special cases, the Ricci
flow in Lorentzian spacetime does indeed exist.

The Ricci flow system in a Lorentzian spacetime, as a hyperbolic-parabolic
competing system, indeed has a similar counterpart in real physical
systems, namely the heat conduction equation with a finite heat conduction
velocity, known as the Cattaneo equation (see e.g. \citep{1989Heat}).
In this context, the finite heat conduction velocity plays a role
analogous to the finite upper limit of the speed of light in Lorentzian
spacetime, while the dissipative structure of heat conduction mimics
the role of the Ricci flow (although the Ricci flow describes a coarse-graining
process in terms of renormalization scale, this differs from the Cattaneo
equation, which describes a dissipative process in time). It is known
that such systems are well-posed, and when the dissipative structure
dominates the competition, the system possesses a monotonic entropy
functional, similar to the case of the Fourier heat conduction equation
with infinite heat conduction velocity in conventional 3-space, which
also admits a monotonic entropy functional.

Physically, if spatial coordinates gradually broaden and become dominated
by long wavelengths under the influence of the Ricci flow, then, due
to the constancy of the speed of light, the width of the temporal
coordinate will gradually narrow and become dominated by high frequencies.
In other words, as spatial coordinates gradually become blurred and
lose small-scale information, the temporal coordinate becomes increasingly
precise, which seemingly implies the acquisition of additional information---this
is superficially why backward-parabolic equations are ill-posed. However,
from the perspective of quantum reference frames, when the frame fields
that measure time and space, along with the probability density $u$
of the frame fields, form a coupled system, this scenario is not impossible.
Because the normalized density u suppresses the probability of high-frequency
modes occurring in the clock frame field, so although the temporal
coordinate becomes increasingly precise and its energy spectrum broadens,
the high-frequency portion of the spectrum is suppressed, ultimately
broadening into a maximum-entropy blackbody spectrum rather than being
completely dominated by high frequencies. This is analogous to the
``ultraviolet catastrophe'' problem in blackbody radiation, where
the probability of high-frequency transitions between energy levels
of harmonic oscillators is suppressed, preventing the energy spectrum
from diverging in the ultraviolet. Moreover, the information lost
in the spacelike modes can be greater than the information gained
in the timelike modes, and the entropy of the entire system, as reflected
by the probability density $u$ of the frame fields, still increases.
The 4-spacetime line element continues to irreversibly blur, and no
additional information is introduced into the entire system. These
observations suggest that we should seek a monotonically varying entropy-like
functional constructed from the probability density of the frame fields
to control the Ricci flow of the entire Lorentzian curved spacetime.
As discovered by Perelman, in a three-dimensional Riemannian space
with density $(M^{3},g_{ij},u)$, there exist monotonically varying
entropy functionals constructed from the three-dimensional metric
$g_{ij}$ and the geometric density $u$ , such that the Ricci flow
(strictly speaking, the Ricci-DeTurck flow) and the conjugate heat
flow of the coupled $u$ density are gradient flows of this entropy
functional, in the language of physics, it means that the flow equations
are derived from the variations of the entropy functional. Consequently,
even if the conjugate heat flow of the $u$ density is backward-parabolic
in Euclidean space, solutions still exist. Furthermore, when the Ricci
flow gradually tends to develop singularities locally, the finiteness
of these monotonically varying functionals in three-dimensional space
ensures that curvature blow-up does not occur under finite-scale changes
(no local collapsing).

Therefore, the aim of this paper is to construct monotonically varying
entropy functionals for four-dimensional Lorentzian curved manifolds
under some proper physical boundary conditions, analogous to those
constructed by Perelman for three-dimensional Euclidean Riemannian
manifolds. When high-frequency blow-up occurs in the timelike modes,
it would lead to the divergence of such functionals, which contradicts
the conclusion that these monotonically varying functionals exhibit
finite, monotonic changes under finite flow parameters. Consequently,
this can be used to prove, by contradiction, the well-posedness of
the timelike modes in the Ricci flow, demonstrating that the Ricci
flow can be effectively applied to Lorentzian spacetime.

The existence of monotonic functionals in the renormalization of quantum
field theory has long been one of the fundamental questions in mathematical
physics. Here we'll conduct a somewhat incomplete review on this issue.
Zamolodchikov \citep{1986Irreversibility} proved that the central
charge $c$ of a $d=2$ conformal field theory is monotonically non-decreasing
under the renormalization group flow from the ultraviolet to the infrared
regime. The $c$-function behaves like an entropy, effectively counting
the number of degrees of freedom in the system during the renormalization
process. For the case of quantum field theories in real four-dimensional
spacetime (e.g. \citep{1989Derivation,1990Analogs,PhysRevD.54.5163}),
Cardy \citep{1988Is} proposed that the expectation value of the trace
of the energy-momentum tensor in four-dimensional theories could serve
as a generalization of the two-dimensional $c$-function. Proposing
the ADM energy as a monotonic quantity is given in \citep{Gutperle:2002ki}.
Entanglement entropy is suggested as an alternative version of the
c-theorem for (1+1)-dimensional QFT in \citep{Casini:2004bw}. The
article \citep{2005Irreversibility} discusses the generalization
of the c-function to world-sheet RG flow on noncompact spacetimes
by using Perelman-type functionals, which is similar with our work
in some aspects, but the monotonic functionals we obtain do not depend
on whether the background scalar curvature is positive or negative
which is different from their discussions. \citep{Oliynyk:2005ak,Tseytlin:2006ak}
also generalize Perelman's functionals to discuss the RG flow in sigma
models. Komargodski and Schwimmer suggested the existence of a monotonic
$a$-function in $d=4$ quantum field theories as a generalization
for four-dimensional cases \citep{2011On}. Some literature \citep{Ruchin:2013azz,Bubuianu:2019hfh,Bubuianu:2019yfr,Vacaru:2019lpl}
has explored extending Perelman's monotonic entropy functional to
relativistic forms in 3+1 dimensional spacetime. 

From the perspective of quantum reference frames, which can be viewed
as a special type of $d=4-\epsilon$ quantum fields theory, more precisely,
a nonlinear sigma model, the construction of entropy generalized to
four-dimensional spacetime in this paper is equivalent to providing
a construction of a monotonic functional for a nonlinear sigma model
where the target spacetime is a four-dimensional Lorentzian spacetime
and the base space is $d=4-\epsilon$ dimensional. Based on the quantum
equivalence principle \citep{2024arXiv240809630L}, which states that
the properties of the frame field (the non-dynamical part), such as
the average values of field quantities and their second-moment fluctuations,
measure universal properties of spacetime (average values of spacetime
coordinates, second-moment broadening of coordinates, etc.). Therefore,
this monotonic relative entropy functional also measures the entropy
of the Lorentzian spacetime itself. The existence of the global control
of the Lorentzian spacetime and gravity of the quantum version by
using the monotonic functionals is profoundly significant.

This paper is structured as follows: In Section II, we will provide
a more detailed explanation of Perelman's monotonic entropy functional
in three-dimensional Riemannian spaces, as well as its application
to the Ricci flow in such spaces. In Section III, we attempt to generalize
Perelman's monotonic entropy functional from three-dimensional Riemannian
spaces to four-dimensional Lorentzian curved spacetime. By appropriately
defining the $u$ density in four-dimensional Lorentzian spacetime,
we construct the Shannon entropy and its relative entropy, and prove
the monotonicity of the relative entropy (which we refer to as the
H-theorem for spacetime). Starting from the Shannon entropy, we derive
a generalized monotonic entropy functional in four-dimensional Lorentzian
spacetime, analogous to Perelman's functional in three-dimensional
spaces. We demonstrate that the gradient flow of these monotonic entropy
functionals yields the Ricci flow equations for four-dimensional Lorentzian
spacetime, along with the conjugate heat flow equations for the coupled
$u$ density. In Section IV, we provide several examples to illustrate
the physical significance and applications of our monotonic entropy
functional in gravitational systems. Finally, in the Section V, we
summarize the paper.

\section{Perelman's Entropic Formula in 3D Riemannian Manifolds}

Perelman's approach to handling the Ricci flow of three-dimensional
Riemannian manifolds $(M^{3},g_{ij})$ involves considering a three-dimensional
Riemannian manifold $(M^{3},g_{ij},u)$ endowed with a density function
$u$, where the density $u(X)$ satisfies a normalization condition
\begin{equation}
\int_{M^{3}}d^{3}X\sqrt{g(X)}u(X)=1\label{eq:3d u normalization}
\end{equation}
where $X$ represent the coordinates in three-dimensional Riemannian
geometry, and $\sqrt{g(X)}$ is the volume element of the three-dimensional
Riemannian geometry. Now, the metric $g_{ij}$ satisfies the generalized
Ricci-DeTurck equation in the context of the density geometry
\begin{equation}
\frac{\partial g_{ij}}{\partial t}=-2\left(R_{ij}-\nabla_{i}\nabla_{j}\log u\right)\label{eq:Ricci-DeTurck flow}
\end{equation}
in which $t$ is the parameter of the Ricci flow, and $\nabla_{i}\nabla_{j}\log u$
represents a diffeomorphic transformation performed on the Ricci curvature
$R_{ij}$. Therefore, this equation is equivalent to the original
Ricci flow equation up to a diffeomorphism. Mathematically, the reason
for adding this term is that the principal symbol on the right-hand
side of the original Ricci flow equation (\ref{eq:3d Ricci flow})
is merely non-negative (weakly parabolic). After incorporating this
term, the principal symbol on the right-hand side becomes positive-definite,
transforming the equation into a strongly parabolic one. This simplifies
the analysis of the existence and uniqueness of solutions, the technique
is known as the DeTurck trick.

Because $u$ is defined by a normalization constraint (\ref{eq:3d u normalization}),
the constraint remains invariant under the evolution of the Ricci
flow
\begin{equation}
\frac{\partial}{\partial t}\left(u\sqrt{g}\right)=0\label{eq:3d u normalized}
\end{equation}
Since the volume element $\sqrt{g}$ evolves under the Ricci flow,
we obtain the flow equation for the $u$ density
\begin{align}
\sqrt{g}\frac{\partial u}{\partial t}+u\frac{\partial}{\partial t}\sqrt{g} & =0\nonumber \\
\frac{\partial u}{\partial t} & =-\frac{u}{\sqrt{g}}\frac{\partial\sqrt{g}}{\partial g_{ij}}\frac{\partial g_{ij}}{\partial t}\nonumber \\
 & =\frac{u}{\sqrt{g}}\frac{1}{2}\sqrt{g}g^{ij}2\left(R_{ij}-\nabla_{i}\nabla_{j}\log u\right)\nonumber \\
 & =u\left(R-\frac{\Delta u}{u}\right)\nonumber \\
 & =\left(-\Delta+R\right)u\label{eq:3d u equ}
\end{align}
in which, $\Delta=g^{ij}\nabla_{i}\nabla_{j}$ represents the Laplacian-Beltrami
operator in three-dimensional Riemannian geometry. This equation is
known as the conjugate heat equation for the $u$ density. Note the
negative sign preceding the $\Delta$ operator, as a result, the conjugate
heat equation is backward parabolic with respect to the flow parameter
$t$. Generally, solutions to such equations exhibit unstable high-frequency
blow-up, rendering the problem ill-posed. However, Perelman discovered
that this coupled system of the Ricci-DeTurck equation and the conjugate
heat equation
\begin{equation}
\begin{cases}
\frac{\partial g_{ij}}{\partial t}=-2\left(R_{ij}-\nabla_{i}\nabla_{j}\log u\right)\\
\frac{\partial u}{\partial\tau}=\left(\Delta-R\right)u\\
\frac{d\tau}{dt}=-1
\end{cases}\label{eq:3d coupled equs}
\end{equation}
can be regarded as the gradient flow of a three-dimensional F-functional,
which is composed of the metric $g_{ij}$ and the density $u\equiv e^{-f}$
\begin{equation}
\mathcal{F}_{3}(g,f,\tau)=\int_{M}d^{3}X\sqrt{g}e^{-f}\left(R+|\nabla f|^{2}\right)\label{eq:3d F}
\end{equation}
in which $\tau$ is a backwards flow parameter $d\tau=-dt$.

Because (\ref{eq:3d u normalized}), we have
\begin{equation}
\frac{d\mathcal{F}_{3}}{dt}=-\int_{M^{3}}d^{3}X\sqrt{g}e^{-f}\left[\frac{\partial g_{ij}}{\partial t}\left(R^{ij}+\nabla^{i}\nabla^{j}f\right)\right]
\end{equation}
or
\begin{equation}
\delta\mathcal{F}_{3}=-\int_{M^{3}}d^{3}X\sqrt{g}e^{-f}\left(R^{ij}+\nabla^{i}\nabla^{j}f\right)\delta g_{ij}
\end{equation}
so the Ricci-DeTurck flow (\ref{eq:Ricci-DeTurck flow}) is just a
gradient flow of the F-functional.

From the above results, it is readily apparent that the F-functional
is monotonically non-decreasing with respect to the parameter $t$
\begin{equation}
\frac{d\mathcal{F}_{3}}{dt}=2\int_{M^{3}}d^{3}X\sqrt{g}e^{-f}\left(R_{ij}+\nabla_{i}\nabla_{j}f\right)\left(R^{ij}+\nabla^{i}\nabla^{j}f\right)=2\int_{M^{3}}d^{3}X\sqrt{g}e^{-f}\left|R_{ij}+\nabla_{i}\nabla_{j}f\right|^{2}\ge0\label{eq:3d perfect square}
\end{equation}

Therefore, under a finite flow of the parameter $t$, the metric $g$
will undergo a finite and monotonic change during this process. This
is because if local curvature blow-up occurs in $g$ or if the high-frequency
modes experience exponential growth due to the backwards parabolic
nature of the density $u=e^{-f}$, it would result in an extremely
large curvature term or gradient term $|\nabla f|$ in the F-functional,
causing the integrated F-functional to tend toward infinity. This
contradicts the fact that the F-functional undergoes finite and monotonic
changes within a finite $t$. Hence, this contradiction proves that,
despite the seemingly backwards parabolic nature of the equation for
the $u$ density, its solutions do not exhibit unstable high-frequency
blow-ups. Moreover, precisely because the $u$ density does not blow
up locally, the local volume remains non-collapsing according to the
constraint condition (\ref{eq:3d u normalization}).

A more rigorous local non-collapsing theorem requires a generalizing
the F-functional to a scale-invariant form (invariant under simultaneous
rescaling of $\tau$ and $g$), ensuring that the manifold does not
develop singularities locally under rescaling. This is the W-entropy
functional introduced also by Perelman
\begin{equation}
\mathcal{W}_{3}(g,f,\tau)=\int_{M^{3}}d^{3}X\sqrt{g}u\left[\tau\left(R+|\nabla f|^{2}\right)+f-3\right]\label{eq:3d W}
\end{equation}
in which $u=(4\pi\tau)^{-3/2}e^{-f}$. By using (\ref{eq:3d u normalized})
we also have
\begin{equation}
\frac{\partial\mathcal{W}_{3}}{\partial t}=\int_{M^{3}}d^{3}X\sqrt{g}u\left[\frac{\partial\tau}{\partial t}\left(R+|\nabla f|^{2}\right)+\frac{\partial f}{\partial t}-\tau\frac{\partial g_{ij}}{\partial t}\left(R^{ij}+\nabla^{i}\nabla^{j}f\right)\right]
\end{equation}
so
\begin{equation}
\delta\mathcal{W}_{3}=-\int_{M^{3}}d^{3}X\sqrt{g}u\left[\tau\left(R^{ij}+\nabla^{i}\nabla^{j}f\right)\right]\delta g_{ij}
\end{equation}
That is, the gradient flow of the W-entropy functional remains equivalent
to the Ricci flow, up to diffeomorphisms and rescaling. Furthermore
\begin{align}
\frac{\partial\mathcal{W}_{3}}{\partial t} & =\int_{M^{3}}d^{3}X\sqrt{g}u\left[\frac{\partial\tau}{\partial t}\left(R+|\nabla f|^{2}\right)+\frac{\partial f}{\partial t}-\tau\frac{\partial g_{ij}}{\partial t}\left(R^{ij}+\nabla^{i}\nabla^{j}f\right)\right]\nonumber \\
 & =\int_{M^{3}}d^{3}X\sqrt{g}u\left[-\left(R+|\nabla f|^{2}\right)+\left(-\Delta f-R+\frac{3}{2\tau}\right)+2\tau\left|R_{ij}+\nabla_{i}\nabla_{j}f\right|^{2}\right]\nonumber \\
 & =\int_{M^{3}}d^{3}X\sqrt{g}u\left[-2\left(R+\Delta f\right)+\frac{3}{2\tau}+2\tau\left|R_{ij}+\nabla_{i}\nabla_{j}f\right|^{2}\right]\nonumber \\
 & =2\tau\int_{M^{3}}d^{3}X\sqrt{g}u\left|R_{ij}+\nabla_{i}\nabla_{j}f-\frac{1}{2\tau}g_{ij}\right|^{2}\ge0
\end{align}
Therefore, the W-entropy functional is also monotonically non-decreasing
along the Ricci flow, the equality holds when the three-dimensional
Riemannian manifold $g$ satisfies the Gradient Shrinking Ricci Soliton
(GSRS) equation
\begin{equation}
R_{ij}+\nabla_{i}\nabla_{j}f-\frac{1}{2\tau}g_{ij}=0
\end{equation}
when the W-entropy functional attains its extremum. In addition, for
a 3d Riemannian space, the W-entropy formally as a log-Sobolev inequality
ensures a global boundedness of the W-entropy including its initial
value. Therefore, under a finite flow of the parameter $t$, the W-entropy
functional undergoes only finite changes w.r.t. the boundedness initial
value. This similarly rules out the occurrence of blow-ups in either
the curvature or the $u$ density during the finite-$t$ flow process;
otherwise, the W-entropy functional would diverge during this process.
The existence of these monotonic functionals indicates that the system
of equations (\ref{eq:3d coupled equs}) is well-posed, despite the
fact that some of the equations within the system are backwards parabolic.

\section{Generalized Entropic Formula for 4D Lorentzian Spacetime}

If we regard the Ricci flow as a dynamical system, then its first
integrals, such as conserved quantities (e.g., energy) or monotonic
quantities (e.g., entropy), exert significant constraints on the behavior
of the system. Therefore, analogous to the previous section, to demonstrate
that even though part of the equations in Lorentzian spacetime (specifically,
those governing the timelike modes of the metric) are backwards parabolic,
ill-defined issues like high-frequency blow-ups will not occur, we
need to identify similar monotonic functionals in Lorentzian spacetime.
These functionals should ensure that the Ricci flow equations in Lorentzian
spacetime serve as gradient flows of some monotonic functionals.

\subsection{Shannon Entropy and Generalized F-functional}

In four-dimensional Lorentzian spacetime, since the timelike metric
component are negative-definite, to preserve the positive-definiteness
of the Lorentzian 4-volume and the $u$ 4-density, we will generalize
the definition of the $u$ density to a Lorentzian spacetime with
density $(M^{3+1},g_{\mu\nu},u)$ 
\begin{equation}
\int_{M^{D}}d^{D}X\sqrt{\left|g\right|}u\equiv1\label{eq:4d u normalization}
\end{equation}
in which the determinant of the Lorentzian spacetime metric is taken
with an absolute value symbol, ensuring that the density $u$ remains
positive-definite without introducing any unwanted explicit imaginary
$i$ factor. And we will set $D=4$ throughout the paper, to demonstrate
the universality of the results. In the subsequent discussion, the
meaning of the spacetime integral in four-dimensional Lorentzian spacetime
is defined as 
\begin{equation}
\int_{M^{D}}d^{D}X\equiv\int_{T_{1}}^{T_{2}}dT\int_{\Sigma_{T}}d^{D-1}X
\end{equation}
where the spacelike hypersurfaces $\Sigma_{T_{1}}$ and $\Sigma_{T_{2}}$
at the initial time $T_{1}$ and the final time $T_{2}$, respectively,
are considered asymptotically fixed. 

When the space and time are put on an equal footing, at the spacetime
infinity we consider the variation of the metric satisfy 
\begin{equation}
\left.\delta g_{\mu\nu}\right|_{\partial M^{D}}=0\label{eq:g boundary}
\end{equation}
Such boundary setting of the Lorentzian spacetime are not only physically
natural but also ensure that no additional boundary terms arise under
the variational principles, so it is widely used in general relativity.
The conservation of density in a positive-defined spacetime volume
(\ref{eq:4d u normalization}) ensures that the probability current
falls off at spacetime infinity, satisfying a physical boundary condition
\begin{equation}
\left.J_{\mu}\right|_{\partial M^{D}}=\left.\nabla_{\mu}u\right|_{\partial M^{D}}=0\label{eq:u boudary}
\end{equation}
An alternative choice is the no-boundary condition proposed in the
context of cosmology, in this scenario, it gives a more direct analog
to generalize the three-dimensional compact case to the four-dimensional
Lorentzian spacetime case. In the following discussions, we consider
(\ref{eq:g boundary}) and (\ref{eq:u boudary}) as physically natural
spacetime boundary conditions under which the generalized functionals
in four-dimensional Lorentzian spacetime have more concise forms without
extra boundary terms, so that they can be formally comparable to the
ones in the compact Riemannian manifolds.

Under this definition of the $u$ density in four-dimensional Lorentz
spacetime, we can similarly derive the conjugate heat equation for
$u$ based on the condition $\frac{\partial}{\partial t}\left(u\sqrt{|g|}\right)=0$
\begin{align}
\sqrt{|g|}\frac{\partial u}{\partial t}+u\frac{\partial}{\partial t}\sqrt{|g|} & =0\nonumber \\
\frac{\partial u}{\partial t} & =-\frac{u}{\sqrt{|g|}}\frac{\partial\sqrt{|g|}}{\partial g_{\mu\nu}}\frac{\partial g_{\mu\nu}}{\partial t}\nonumber \\
 & =\frac{u}{\sqrt{|g|}}\frac{1}{2}\sqrt{|g|}g^{\mu\nu}2\left(R_{\mu\nu}-\nabla_{\mu}\nabla_{\nu}\log u\right)\nonumber \\
 & =u\left(R-\frac{\boxempty u}{u}\right)\nonumber \\
 & =\left(-\boxempty+R\right)u\label{eq:4d u equ}
\end{align}
which is generalization of (\ref{eq:3d u equ}). In the equation,
$\nabla_{\mu}$ is a covariant derivative, and $\boxempty=g^{\mu\nu}\nabla_{\mu}\nabla_{\nu}$
is the Laplacian-Beltrami operator in four-dimensional Lorentzian
spacetime, and also in which we also have used the Ricci-DeTurck flow
equation in the Lorentzian spacetime
\begin{equation}
\frac{\partial g_{\mu\nu}}{\partial t}=-2\left(R_{\mu\nu}-\nabla_{\mu}\nabla_{\nu}\log u\right)\label{eq:4d Ricci-DeTurck}
\end{equation}
Thus, we aim to identify certain monotonic functionals such that their
gradient flows yield a system of equations analogous to (\ref{eq:3d coupled equs})
in four-dimensional Lorentzian spacetime
\begin{equation}
\begin{cases}
\frac{\partial g_{\mu\nu}}{\partial t}=-2\left(R_{\mu\nu}-\nabla_{\mu}\nabla_{\nu}\log u\right)\\
\frac{\partial u}{\partial\tau}=\left(\boxempty-R\right)u\\
\frac{d\tau}{dt}=-1
\end{cases}\label{eq:4d coupled equs}
\end{equation}

Here, we still employ a backwards flow $\tau$ parameter, analogous
to that used in the Riemannian settings, for the sake of convenience
when formulating the conjugate heat equation. This approach ensures
that the resulting system of equations maintains a formal similarity
to those in Riemannian case. That is, if we Wick rotate the equation,
it becomes a four dimensional Euclidean version, whose fundamental
solution is easy to be written down, and hence it is more transparent
to obtain a formal fundamental solution of the Lorentzian version
by a Wick rotating back. The formal fundamental solution also formally
exhibits the ``backwards parabolic'' or ``high-frequency blow-up''
problem in its specific modes. The ``blow up'' appears to violate
the constraint (\ref{eq:4d u normalization}), but bear in mind that
the conjugate heat equation originates from this very constraint.
Therefore, this problem must merely be an illusion. The solution of
the well-posedness problem of the conjugate heat equation also relies
on the existence of a global control of the whole coupled system of
$u$ and $g_{\mu\nu}$ by certain monotonic functionals, rather than
focusing solely on the isolated $u$ or $g_{\mu\nu}$ system, the
same solution in the three-dimensional Riemannian case. 

In the theory of quantum reference frames, the forward flow parameter
$t$ is interpreted as being dependent on the square of the truncated
momentum of the frame fields \citep{Luo2014The,Luo2015Dark,Luo:2015pca,Luo:2019iby,Luo:2021zpi,Luo:2022goc,Luo:2022statistics,Luo:2022ywl,Luo:2023eqf,2023AnPhy.45869452L,2024arXiv240809630L}.
The process where $t$ flows from $+\infty$ to $0$ can be viewed
as the frame fields evolving from the ultraviolet (short-distance)
regime to the infrared (long-distance) regime, gradually averaging
out the short-distance degrees of freedom. The $u$ density represents
the ensemble probability density of the 4-spacetime frame fields and
is also equivalent to the ratio or Jacobian between the local volume
and the standard laboratory volume.

Since the definition of the four-dimensional Lorentzian geometric
$u$ density has now been generalized to a positive-definite form
(\ref{eq:4d u normalization}), we can utilize this positive-definite
$u$ density to define a standard Shannon entropy for the Lorentzian
spacetime
\begin{equation}
N(M^{D})=-\int_{M^{D}}d^{D}X\sqrt{|g|}u\log u
\end{equation}

It can be proven that the derivative of this Shannon entropy with
respect to $\tau$ yields a generalized form of Perelman's F-functional
(\ref{eq:3d F}) for three-dimensional Riemannian manifolds, adapted
to the context of four-dimensional Lorentzian spacetime
\begin{align}
\frac{dN}{d\tau} & =-\frac{d}{d\tau}\int_{M^{D}}d^{D}X\sqrt{|g|}u\log u\nonumber \\
 & =-\int_{M^{D}}d^{D}X\sqrt{|g|}\left[(1+\log u)\frac{\partial u}{\partial\tau}+Ru\log u\right]\nonumber \\
 & =-\int_{M^{D}}d^{D}X\sqrt{|g|}\left[(1+\log u)\left(\boxempty u-Ru\right)+Ru\log u\right]\nonumber \\
 & =\int_{M^{D}}d^{D}X\sqrt{|g|}\left(\frac{(\nabla u)^{2}}{u}+Ru\right)\nonumber \\
 & =\int_{M^{D}}d^{D}X\sqrt{|g|}u\left(R+(\nabla f)^{2}\right)=\mathcal{F}\label{eq:4d F}
\end{align}
In the second line of the derivation, we have used the Ricci flow
of the volume element
\begin{equation}
\frac{d}{d\tau}\left(\sqrt{|g|}d^{D}X\right)=R\left(\sqrt{|g|}d^{D}X\right)
\end{equation}
In the third line, we have use the boundary condition (\ref{eq:u boudary}),
i.e. $\int_{M^{D}}d^{D}X\sqrt{|g|}\boxempty u=\int_{\partial M^{D}}dS^{\mu}J_{\mu}=0$,
so
\begin{equation}
-\int_{M^{D}}d^{D}X\sqrt{|g|}\left(\boxempty u\right)\log u=\int_{M^{D}}d^{D}X\sqrt{|g|}u\left(\boxempty\log u\right)=\int_{M^{D}}d^{D}X\sqrt{|g|}\left[\frac{\left(\nabla u\right)^{2}}{u}-\boxempty u\right]=\int_{M^{D}}d^{D}X\sqrt{|g|}\frac{\left(\nabla u\right)^{2}}{u}
\end{equation}
where $(\nabla u)^{2}\equiv g^{\mu\nu}\nabla_{\mu}u\nabla_{\nu}u$. 

The proof of the positivity of the F-functional (\ref{eq:4d F}) is
equivalent to the proof the positive energy theorem of general relativity
at the Ricci flow level. To find whether the F-functional is positive
along the flow, we need to study its monotonicity, the derivative
of the F-functional in four-dimensional Lorentzian spacetime turns
out to be a perfect square
\begin{equation}
\frac{d\mathcal{F}}{dt}=-\int_{M^{D}}d^{D}X\sqrt{|g|}u\frac{\partial g_{\mu\nu}}{\partial t}\left(R^{\mu\nu}+\nabla^{\mu}\nabla^{\nu}f\right)=2\int_{M^{D}}d^{D}X\sqrt{|g|}u\left(R_{\mu\nu}+\nabla_{\mu}\nabla_{\nu}f\right)^{2}\label{eq:dF/dt>=00003D0}
\end{equation}
Here the square of the Bakry-\'{E}mery curvature $R_{\mu\nu}+\nabla_{\mu}\nabla_{\nu}f$
refers to performing a trace contraction after raising and lowering
indices using the metric
\begin{equation}
\left(R_{\mu\nu}+\nabla_{\mu}\nabla_{\nu}f\right)^{2}=g^{\mu\rho}g^{\nu\sigma}\left(R_{\mu\nu}+\nabla_{\mu}\nabla_{\nu}f\right)\left(R_{\rho\sigma}+\nabla_{\rho}\nabla_{\sigma}f\right)=\left(R_{\nu}^{\mu}+\nabla^{\mu}\nabla_{\nu}f\right)\left(R_{\mu}^{\nu}+\nabla_{\mu}\nabla^{\nu}f\right)\label{eq:perfect square}
\end{equation}

But unfortunately, the perfect squared quantity is indefinite in general
for Lorentz signature, because the Bakry-\'{E}mery curvature $R_{\mu\nu}+\nabla_{\mu}\nabla_{\nu}f$
is generally not self-adjoint under the Lorentz metric, so its eigenvalues
are complex in general, so different from the 3d Riemannian case,
it is not globally non-negative, which is a main difficulty to apply
Perelman's functionals for Lorentz manifolds. 

However, thanks to the DeTurck's trick, one can observe that the Bakry-\'{E}mery
curvature exhibits enough gauge arbitrariness to modulate its spectrum,
which is modulated by the DeTurck term, i.e. the Hessian $\nabla_{\mu}\nabla_{\nu}f$.
Note that the non-self-adjointness mainly comes from the Ricci curvature,
while the Hessian is self-adjoint under the given boundary condition
(\ref{eq:u boudary}), so the lower bound of the real part of eigenvalues
of the Bakry-\'{E}mery curvature is controlled by the Hessian. And
also because the Hessian is of order $\sim O(g_{\mu\nu})$ while the
curvature is of higher order $\sim O(\partial^{2}g_{\mu\nu})$, it
indicates that the long-flow-time behavior of the system is inevitably
determined by the real part coming from the control of the Hessian.
Therefore, if the curvature is initially bounded, the gauge arbitrariness
allows us to select a sufficiently ``steep'' initial Hessian of
$f$ to set a sufficiently high lower bound for the real part, thereby
suppressing the imaginary part arising from its non-self-adjointness
in the Lorentz signature. In other words, the choice of the gauge
allows us to stabilize the system via a sufficient initial Hessian,
suppressing the oscillation due to the imaginary part, such that the
system behaves as a strongly damped stablized system. By employing
a strong enough convex function $f$ as a gauge choice, we can enable
the flow to persist for a sufficiently long flow-time, and because
the blow-up time does not depend directly on the gauge choice, so
it can even persist to the maximum existing time when local curvature
may blow-up and hence the Hessian term loses its dominance. In fact,
if the monotonicity of the flow valid, near the blow-up time, the
geometry gradually resembles a gradient Ricci soliton $R_{\mu\nu}+\nabla_{\mu}\nabla_{\nu}f=\lambda g_{\mu\nu}$,
leaving only a real eigenvalue $\lambda\in\mathbb{R}$ as the soliton
parameter while the imaginary parts of the eigenvalues become completely
decoupled, the real-part dominance automatically satisfies, the effectiveness
of the real-part dominance coming from the gauge choice can be even
valid to the maximum existing time of the flow. In this sense, although,
under the Lorentzian signature, it is no longer possible to maintain
the global non-negativity of the perfect square term as in the 3d
Riemannian space as (\ref{eq:3d perfect square}), it is still feasible
to keep the perfect square term (\ref{eq:perfect square}) non-negative
for a sufficiently long period of time (semi-globally). In this case,
(\ref{eq:dF/dt>=00003D0}) can be non-negative for sufficient long
time, although it is weaker than the 3d Riemannian case, it still
indicates that the generalized Perelman's F-functional can be applied
in a physical Lorentzian spacetime, it is monotonically non-decreasing
under the constraint (\ref{eq:4d u normalization}), which gives a
``semi-global'' control to the Ricci-DeTurck flow of the Lorentz
spacetime (compared to the ``global'' control in the Riemannian
case). Therefore, the DeTurck trick not only modifies the original
Lorentzian Ricci flow (without the Hessian) from weakly hyperbolic
to strongly hyperbolic, but also stablizes the flow by making the
real part dominants over the imaginary part of the spectra. Although
the DeTurck gauge fixing term cannot fully eliminate the oscillations
caused by the imaginary parts of eigenvalues or the time-evolution
degrees of freedom, thereby transforming the spacetime into a completely
static one (the Ricci flow of a static Lorentzian spacetime \citep{2024Short},
as a semi-Riemannian one, is parabolic, well-posed, and has monotonic
entropy), the gauge fixing does eliminate those non-physical degrees
of freedom. As a result, the entropy of the remaining physical degrees
of freedom in the non-static Lorentzian spacetime is also monotonic.
Although, under the hyperbolic equations in Lorentzian spacetime,
we lack the global control provided by the maximum principle for parabolic
equations, if the F-functional (\ref{eq:4d F}) is positive initially,
its monotonicity ensures it can keep positive for long flow-time,
which, in certain sense, is a positive energy theorem of the Ricci
flow or quantum version.

\subsection{Relative Entropy and H-theorem}

During the Ricci flow process in four-dimensional Lorentzian spacetime,
equilibrium state may develop locally where the entropy gets extremal
value, similar to the case in three-dimensional Riemannian manifolds.
Near such equilibrium scale $t_{*}$, where the backwards flow parameter
$\tau=t_{*}-t\rightarrow0$, since the long-flow-time behavior of
the system is controlled by the dissipative real-part of the Bakry-\'{E}mery
curvature as mentioned above, the initial condition of the conjugate
heat equation can be given by a Gaussian-type density (up to a gauge
choice)
\begin{equation}
u_{*}(X)=\frac{1}{(4\pi\tau)^{D/2}}\exp\left(-\frac{\left|X^{2}\right|}{4\tau}\right),\quad(\tau\rightarrow0)\label{eq:initial density}
\end{equation}
We define $\left|X^{2}\right|=\left|g_{\mu\nu}X^{\mu}X^{\nu}\right|$
being a Lorentzian distance with an absolute symbol. Such an absolute
symbol ensures that the initial condition of $u$ falls off both in
spatial and temporal directions, and hence satisfying the constraint
(\ref{eq:4d u normalization}) and the boundary condition (\ref{eq:u boudary}).
The non-differentiability happens at the zero point of the absolute
symbol, i.e. the light cone, which profoundly reflects the discontinuity
in both sides of the light cone and the singular nature of the light
cone.

It is worth stressing that the initial density is not a distribution
concentrated at a single point, but rather a uniform distribution
concentrated on the light cone 3-hypersurface, which is locally a
3-Euclidean space as the 3-hypersurface of the 4-spacetime. When the
uniform distribution in the 3-space evolves uniformly in time direction
(i.e. a uniform distribution in a flat spacetime $\mathbb{R}^{D}$)
it gives a maximal entropy, in analogous to an equilibrium state
\begin{equation}
N_{*}=-\int_{\mathbb{R}^{D}}d^{D}X\sqrt{|g|}u_{*}\log u_{*}=\frac{D}{2}\left[1+\log(4\pi\tau)\right]
\end{equation}
We can define the relative entropy $\tilde{N}$ of the Shannon entropy
$N$ in a non-equilibrium state with respect to the equilibrium entropy
$N_{*}$
\begin{align}
\tilde{N} & =N-N_{*}\nonumber \\
 & =\int_{M^{D}}d^{D}X\sqrt{|g|}u\left[f+\frac{D}{2}\log(4\pi\tau)\right]-\int_{\mathbb{R}^{D}}d^{D}X\sqrt{|g|}u\frac{D}{2}\left[1+\log(4\pi\tau)\right]\nonumber \\
 & =\int_{M^{D}}d^{D}X\sqrt{|g|}u\left(f-\frac{D}{2}\right)\label{eq:N-tilde}
\end{align}
The relative entropy is also equivalent to defining the Shannon entropy
by using a dimensionless relative density $\tilde{u}=u/u_{*}$. Because
$\left(R_{\mu\nu}-\nabla_{\mu}\nabla_{\nu}\log u\right)^{2}$ can
be non-negative by choosing an appropriate $u$ for a long flow-time,
within the period of time, we can use the Cauchy-Schwarz inequality
and Jensen inequality 
\begin{align}
\frac{d\mathcal{F}}{dt} & =2\int_{M^{D}}d^{D}X\sqrt{|g|}u\left(R_{\mu\nu}-\nabla_{\mu}\nabla_{\nu}\log u\right)^{2}\label{eq:dF/dt}\\
 & \ge\frac{2}{D}\int_{M^{D}}d^{D}X\sqrt{|g|}u\left(R-\boxempty\log u\right)^{2}\nonumber \\
 & \ge\frac{2}{D}\left[\int_{M^{D}}d^{D}X\sqrt{|g|}u\left(R+(\nabla f)^{2}\right)\right]^{2}\nonumber \\
 & =\frac{2}{D}\mathcal{F}^{2}\ge0\nonumber 
\end{align}
in which the equal sign can be achieved only when $\frac{d\mathcal{F}_{*}}{dt}=\frac{2}{D}\mathcal{F}_{*}^{2}$.
So we have the extremal value $\mathcal{F}_{*}$ at the equilibrium
state 
\begin{equation}
\frac{dN_{*}}{d\tau}=\mathcal{F}_{*}=\frac{D}{2\tau}\label{eq:4d F*}
\end{equation}
Therefore, it can be concluded that, the relative entropy $\tilde{N}$
is monotonically non-decreasing during the Ricci flow for sufficient
long time, in any condition of the curvature
\begin{equation}
\frac{d\tilde{N}}{dt}=\frac{dN}{dt}-\frac{dN_{*}}{dt}=-\mathcal{F}+\mathcal{F}_{*}=-\mathcal{F}+\frac{D}{2\tau}\ge0\label{eq:dNtilde/dt>=00003D0}
\end{equation}
under the constraint (\ref{eq:4d u normalization}). The equality
holds only when the Shannon entropy $N$ eventually flows to its extremal
value $N_{*}$, and this is the reason why we call it entropy.

Given that the Gaussian-like of the initial density $u_{*}$ resembles
the Boltzmann-Maxwell distribution in statistical mechanics, the Ricci
flow parameter $t$ is analogous to the Newtonian time, the conjugate
heat equation bears similarity to the Boltzmann equation for a dilute
gas, and the relative entropy is akin to Boltzmann's monotonic H-functional
describing the relaxation of a dilute gas from a non-equilibrium state
to an equilibrium state, we can analogously refer to (\ref{eq:dNtilde/dt>=00003D0})
as the H-theorem for Lorentzian spacetime. Similarly, the monotonicity
of this relative entropy can be viewed as describing the process by
which Lorentzian spacetime evolves under the Ricci flow from a non-equilibrium
state toward an equilibrium state where entropy attains its extremal
value.

\subsection{Generalized W-functional}

It can be observed that the negative of the Legendre transformation
of the relative entropy $\tilde{N}$ with respect to $\tau^{-1}$
yields a functional that is formally analogous to the W-entropy functional
in three-dimensional Riemannian geometry, representing a generalization
of the W-entropy functional to four-dimensional Lorentzian spacetime
\begin{align}
\mathcal{W}(M^{D},g,u,\tau) & \equiv-\left(\tau^{-1}\frac{d\tilde{N}}{d\tau^{-1}}-\tilde{N}\right)=\tau\frac{d\tilde{N}}{d\tau}+\tilde{N}=\tau\tilde{\mathcal{F}}+\tilde{N}\nonumber \\
 & =\int_{M^{D}}d^{D}X\sqrt{|g|}u\left[\tau\left(R+(\nabla f)^{2}\right)-\frac{D}{2}+\left(f-\frac{D}{2}\right)\right]\nonumber \\
 & =\int_{M^{D}}d^{D}X\sqrt{|g|}u\left[\tau\left(R+(\nabla f)^{2}\right)+f-D\right]\label{eq:4d W}
\end{align}
in which we have used the relative F-functional
\begin{equation}
\frac{d\tilde{N}}{d\tau}=\mathcal{F}-\mathcal{F}_{*}=\tilde{\mathcal{F}}=\int_{M^{D}}d^{D}X\sqrt{|g|}u\left[\tau\left(R+(\nabla f)^{2}\right)-\frac{D}{2}\right]\label{eq:4d Ftilde}
\end{equation}
and (\ref{eq:dF/dt}) (\ref{eq:4d F*}) (\ref{eq:4d Ftilde}) are
also used. The W-entropy functional obtained here is exactly the same
in form as the three-dimensional Riemannian one given by Perelman
(\ref{eq:3d W}), and it is also applicable to four-dimensional spacetimes
with a Lorentzian signature. As the Legendre transform of the relative
entropy functional $\tilde{N}$, the W-entropy functional is merely
another measure of the system's entropy.

It can also be proven that, under the constraint (\ref{eq:4d u normalization}),
the W-entropy functional for this four-dimensional Lorentzian spacetime
is monotonically non-decreasing along the flow of the Ricci flow parameter
$t$, no matter the curvature is positive or negative
\begin{align}
\frac{d\mathcal{W}}{dt} & \equiv-\tilde{\mathcal{F}}-\tau\frac{d\tilde{\mathcal{F}}}{d\tau}-\frac{d\tilde{N}}{d\tau}=-\tau\frac{d\tilde{\mathcal{F}}}{d\tau}-2\tilde{\mathcal{F}}=-\tau\left(\frac{d\mathcal{F}}{d\tau}-\frac{d\mathcal{F}_{*}}{d\tau}\right)-2\left(\mathcal{F}-\mathcal{F}_{*}\right)\nonumber \\
 & =\int_{M^{D}}d^{D}X\sqrt{|g|}u\left[2\tau\left(R_{\mu\nu}+\nabla_{\mu}\nabla_{\nu}f\right)^{2}-\frac{D}{2\tau}-2\left(R+(\nabla f)^{2}-\frac{D}{2\tau}\right)\right]\nonumber \\
 & =\int_{M^{D}}d^{D}X\sqrt{|g|}u\left[2\tau\left(R_{\mu\nu}+\nabla_{\mu}\nabla_{\nu}f\right)^{2}-2\left(R+\boxempty f\right)+\frac{D}{2\tau}\right]\nonumber \\
 & =\int_{M^{D}}d^{D}X\sqrt{|g|}u\left[2\tau\left(R_{\mu\nu}+\nabla_{\mu}\nabla_{\nu}f-\frac{1}{2\tau}g_{\mu\nu}\right)^{2}\right]\ge0
\end{align}
in which (\ref{eq:dF/dt}), (\ref{eq:4d F*}) and (\ref{eq:4d Ftilde})
have been used. The derivative of the W-functional also turns out
to be a perfect square which is non-negative even in Lorentzian signature.
The equality holds when the four-dimensional Lorentzian spacetime
satisfies the Gradient Shrinking Ricci Soliton (GSRS) equation
\begin{equation}
R_{\mu\nu}+\nabla_{\mu}\nabla_{\nu}f-\frac{1}{2\tau}g_{\mu\nu}=0\label{eq:GSRS}
\end{equation}

This type of soliton configuration represents a spacetime configuration
where the relative entropy of spacetime attains its extremal value,
and thus, it is typically also a maximally symmetric spacetime configuration.
Such a configuration serves as a generalization of Einstein manifolds.
Under the flow of the Ricci flow parameter, this type of spacetime
configuration does not alter its shape but only changes its size.
Therefore, up to a rescaling, this spacetime configuration constitutes
a fixed-point configuration under the Ricci flow. In the theory of
quantum reference frames, this equation, as a generalization of the
Einstein equation in the infrared limit of gravity, shares the same
formal independence from the metric signature as the Einstein equation,
that is the GSRS equation (\ref{eq:GSRS}) is applicable to both Euclidean
and Lorentzian signature spacetime. It is a crucial equation for studying
long-distance gravitational behavior on cosmic scales and serves as
a model for long-distance cosmic spacetimes. Spacetime configurations
associated with the inflation in the very early universe \citep{2023AnPhy.45869452L},
black hole \citep{Luo:2022statistics}, and accelerated expansion
at late epoch \citep{Luo2014The,Luo2015Dark,Luo:2015pca,Luo:2019iby,Luo:2021zpi,2024arXiv240809630L}
are all related to this type of soliton spacetime configuration.

We have obtained monotonic functionals for four-dimensional Lorentzian
spacetime, namely the F-functional (\ref{eq:4d F}) and the W-entropy
functional (\ref{eq:4d W}). Formally, by setting $D=3$ and replacing
$\sqrt{|g|}$ with $\sqrt{g}$, these functionals reduce to those
introduced by Perelman for three-dimensional compact Riemannian manifolds.

Since the variations
\begin{equation}
\delta\mathcal{F}=-\int_{M^{D}}d^{D}X\sqrt{|g|}u\left(R^{\mu\nu}+\nabla^{\mu}\nabla^{\nu}f\right)\delta g_{\mu\nu}
\end{equation}

\begin{equation}
\delta\mathcal{W}=-\tau\int_{M^{D}}d^{D}X\sqrt{|g|}u\left(R^{\mu\nu}+\nabla^{\mu}\nabla^{\nu}f\right)\delta g_{\mu\nu}
\end{equation}
the gradient flows of these monotonic functionals yield the Ricci-DeTurck
flow (\ref{eq:4d Ricci-DeTurck}) for four-dimensional Lorentzian
spacetime, as well as the conjugate heat flow equation (\ref{eq:4d u equ})
resulting from (\ref{eq:4d Ricci-DeTurck}) and constraint (\ref{eq:4d u normalization}). 

In a completely analogous manner, if high-frequency blow-up occurs
in the $u$ density, then the timelike gradient in $\nabla f$ will
become extremely large in the functionals. Similarly, if high-frequency
blow-up also takes place in the timelike modes of the metric, it will
cause the curvature to become extremely large as well. All of these
scenarios will lead to the divergence of the functionals. Although
there is no Sobolev inequality in Lorentz spacetime to guarantee the
strict boundedness of the initial functional in the mathematical rigor,
but in terms of physical rigor, the non-pathological nature of the
initial spacetime and the boundedness of the initial spacetime functional
can be ensured. Therefore, under the bounded control of these monotonic
functionals, the timelike modes of the metric and $u$ density, which
singly appears to be ``backward parabolic'', as part of the whole
coupled equations, will be well-posed at the level of the physical
rigor, when we set some proper physical backward initial conditions
(i.e. final conditions from the perspective of the forward flow),
up to some gauge choices.

Some examples (e.g. \citep{2020Constructing,Cartas-Fuentevilla:2017cvt})
of Ricci flow for maximally symmetric Lorentzian spacetimes in the
literature essentially depict the Ricci flow evolution from a nearby
non-equilibrium state towards an equilibrium state near the spacetime
configuration with maximum entropy. Since the entropy is already close
to its extremal value, the Ricci flow evolution of these spacetime
configurations can occur spontaneously without requiring much additional
information. Now, the existence of a generally monotonic entropy functional
for Lorentzian spacetime indicates that the Ricci flow for Lorentzian
spacetime still exists even when the initial Lorentzian spacetime
configuration is far from the equilibrium state, i.e. far from maximally
symmetric.

\section{Entropy For Gravity System}

These monotonic functionals in four-dimensional Lorentz spacetime
discussed above are not merely mathematical constructs; they possess
genuine physical meanings in physics. Given the existence of these
monotonic entropy functionals in Lorentz spacetime, these entropies
should play a crucial role in gravitational systems, particularly
in quantum gravity systems (since their origins stem from the quantum
fluctuations of quantum reference frame fields). They serve as global
control and measuring quantities for gravitational and spacetime systems.

\subsection{Shannon Entropy as a Gravity Action}

In classical gravity, gravitation arises from transformations of the
coordinate system (where a specifically chosen non-inertial coordinate
system can eliminate gravity). In the quantum reference frame theory,
quantum gravity also emerges from general (quantum) coordinate transformations.
The relationship between such coordinate transformations and the entropy
of Lorentz spacetime is manifested in the fact that the partition
function of the frame field is not invariant under general quantum
coordinate transformations, a phenomenon known as an anomaly. Considering
the action of the quantum frame fields \citep{Luo2014The,Luo2015Dark,Luo:2015pca,Luo:2019iby,Luo:2021zpi,Luo:2022goc,Luo:2022statistics,Luo:2022ywl,Luo:2023eqf,2023AnPhy.45869452L,2024arXiv240809630L}

\begin{equation}
S[X]=\frac{1}{2}\lambda\int d^{d}x\eta_{ab}g^{\mu\nu}\frac{\partial X_{\mu}}{\partial x_{a}}\frac{\partial X_{\nu}}{\partial x_{b}}\label{eq:NLSM}
\end{equation}
which is a non-linear sigma model in $d=4-\epsilon$ dimensions, $x_{a}$
is the coordinates of the $d=4-\epsilon$ base spacetime or laboratory
frame on which the frame fields $X_{\mu}$ live, $\eta_{ab}$ is the
metric of the base spacetime, without loss of generality, we can adopt
a Euclidean flat metric for the base space, i.e. $\eta_{ab}=\delta_{ab}$,
since the theory is independent of the specific metric and signature
of the base spacetime. $X_{\mu}(x)$ represent $D$ frame fields that
constitute the target spacetime and be promoted as quantum frame fields,
and $g_{\mu\nu}$ is the metric of the target spacetime, which here
is a $D=4$ curved Lorentzian spacetime. $\lambda$ is the coupling
constant, taking the value of the critical density of the universe.

The action $S[X]$ remains invariant under general coordinate transformations
of the spacetime coordinates $X_{\nu}\rightarrow\hat{X}_{\mu}=e_{\mu}^{\nu}X_{\nu}+b_{\mu}$.
However, at the quantum level, the coordinate transformation alters
the functional integral measure of the frame fields 
\begin{align}
\mathscr{D}\hat{X} & \equiv\prod_{x}\prod_{\mu=0}^{3}d\hat{X}_{\mu}(x)\nonumber \\
 & =\prod_{x}\epsilon_{\mu\nu\rho\sigma}e_{\mu}^{0}e_{\nu}^{1}e_{\rho}^{2}e_{\sigma}^{3}dX_{0}(x)dX_{1}(x)dX_{2}(x)dX_{3}(x)\nonumber \\
 & =\prod_{x}\left|\det e_{\mu}^{a}(x)\right|\prod_{x}\prod_{a=0}^{3}dX_{a}(x)\nonumber \\
 & =\left(\prod_{x}\left|\det e_{\mu}^{a}(x)\right|\right)\mathscr{D}X
\end{align}
The Jacobian $|\det e|=\sqrt{|g|}$ of this coordinate transformation
is, in fact, nothing other than the local volume ratio between the
volume and a fiducial volume (e.g. the laboratory frame) or relative
density $\tilde{u}^{-1}$. Therefore, the partition function $Z(M^{D})$
transforms under the coordinate transformations as follows:
\begin{align}
Z(\hat{M}^{D}) & =\int\mathscr{D}\hat{X}\exp\left(-S[\hat{X}]\right)\nonumber \\
 & =\int\left(\prod_{x}|\det e_{\mu}^{a}|\right)\mathscr{D}X\exp\left(-S[X]\right)\nonumber \\
 & =\int\left(\prod_{x}\tilde{u}(\hat{X})^{-1}\right)\mathscr{D}X\exp\left(-S[X]\right)\nonumber \\
 & =\int\left[\prod_{x}e^{-\log\tilde{u}(\hat{X})}\right]\mathscr{D}X\exp\left(-S[X]\right)\nonumber \\
 & =\exp\left(-\lambda\int d^{4}x\log\tilde{u}\right)\int\mathscr{D}X\exp\left(-S[X]\right)\nonumber \\
 & =\exp\left(-\int_{\hat{M}^{D}}d^{D}\hat{X}\sqrt{|g|}u\log\tilde{u}\right)\int\mathscr{D}X\exp\left(-S[X]\right)\nonumber \\
 & =e^{\tilde{N}}Z(M^{D})
\end{align}
in which the volume of the base spacetime is normalized to $\lambda\int d^{4}x=\int_{\hat{M}^{D}}d^{D}\hat{X}\sqrt{|g|}u=1$.
Therefore, the relative entropy $\tilde{N}$ of spacetime measures
the anomaly of general quantum coordinate transformations.

Without loss of generality, we can choose $M^{D}$ to be a classical
laboratory coordinate system that is flat and has no coordinate blurring
or uncertainty, serving as the ultraviolet limit of the frame fields.
In this case, $S[X]=\frac{1}{2}\lambda\int d^{4}xg^{\mu\nu}\partial_{a}x_{\mu}\partial_{a}x_{\nu}=\frac{1}{2}g^{\mu\nu}g_{\mu\nu}=\frac{D}{2}$,
i.e. $Z(M^{D})=e^{-D/2}$ as the partition function for the fiducial
spacetime. The counterterm for the anomaly (ensuring anomaly-free
when returning to the classical laboratory frame) is given by the
difference in relative entropy between the infrared and ultraviolet
regimes
\begin{equation}
\nu=\tilde{N}(\hat{M}_{UV}^{D})-\tilde{N}(\hat{M}_{IR}^{D})=\lim_{\tau\rightarrow\infty}\tilde{N}<0
\end{equation}
Since $\tilde{N}(\hat{M}_{IR}^{D})=0$, so $\nu$ is equivalent to
the ultraviolet limit $\tau\rightarrow\infty$ of the relative entropy.
$\nu$ also contributes a correct cosmological constant $\frac{-2\Lambda}{16\pi G}=\rho_{c}\nu$,
and the ratio $\Omega_{\Lambda}=-\nu=\frac{\rho_{\Lambda}}{\rho_{c}}$
between the ``dark energy density'' $\rho_{\Lambda}$ and the cosmic
critical density $\rho_{c}$ can be determined through purely geometric
methods, since $e^{\nu}<1$ actually represents the asymptotic volume
ratio between the spacetime volume in the infrared flow limit and
its ultraviolet fiducial volume (the laboratory frame). Consequently,
we can obtain the partition function where the anomaly is canceled
in the ultraviolet laboratory frame
\begin{equation}
Z_{cancel}(\hat{M}^{D})=e^{\tilde{N}-\nu-\frac{D}{2}}
\end{equation}
By performing a Schwinger-DeWitt expansion of the relative entropy
with the small parameter $\tau$, we obtain
\begin{align}
\tilde{N}(\hat{M}^{D}) & =\tilde{N}(\hat{M}_{IR}^{D})+\lim_{\tau\rightarrow0}\frac{d\tilde{N}}{d\tau}\tau+O(\tau^{2})\nonumber \\
 & =\lim_{\tau\rightarrow0}\int_{\hat{M}^{D}}d^{D}X\sqrt{|g|}u\;\tau\left[R(0)+(\boxempty f)^{2}-\frac{D}{2\tau}\right]+O(\tau^{2})\nonumber \\
 & \approx\int_{\hat{M}^{D}}d^{D}X\sqrt{|g|}u_{0}R(0)\tau+O(\tau^{2})
\end{align}
in which $R(0)=D(D-1)H_{0}^{2}=12H_{0}^{2}$ represents the scalar
curvature in the infrared limit at $\tau\rightarrow0$, and its value
is equal to 12 times the square of the Hubble parameter $H_{0}$.
In the infrared regime, the $u$ density approaches a constant value
nearly equal to the cosmic critical density, i.e. $u_{0}=\lambda=\rho_{c}=\frac{3H_{0}^{2}}{8\pi G}$.
However, for the purpose of calculating the gradient $\boxempty f$,
we consider an asymptotically equilibrium distribution (\ref{eq:initial density}),
i.e. $f\sim\frac{|X^{2}|}{4\tau}$ at $\tau\rightarrow0$ that depends
on the spacetime coordinates, so we have $\lim_{\tau\rightarrow0}\int d^{D}X\sqrt{|g|}u\,(\boxempty f)^{2}\approx\frac{D}{2\tau}$
which asymptotically cancels the term $\frac{D}{2\tau}$. Finally,
we obtain an action in the infrared (small $\tau$) limit
\begin{align}
-\log Z_{cancel}(\hat{M}^{D}) & =S_{eff}=\int_{M^{D}}d^{D}X\sqrt{|g|}u_{0}\left[\frac{D}{2}-R(0)\tau+\nu+O(R^{2}\tau^{2})\right]\nonumber \\
 & =\int_{M^{D}}d^{D}X\sqrt{|g|}\left[\frac{R(\tau)}{16\pi G}+\lambda\nu+O(R^{2}\tau^{2})\right]
\end{align}
in which we consider that the (backwards) flow of the scalar curvature
$\frac{\partial R}{\partial\tau}=-\boxempty R-2R_{\mu\nu}R^{\mu\nu}$,
therefore, when the infrared curvature is highly uniform and isotropic,
we can assume $\boxempty R(0)=0$ and $R_{\mu\nu}(0)=\frac{1}{D}R(0)g_{\mu\nu}$.
This leads to the solution $R(\tau)=\frac{R(0)}{1+\frac{2}{D}R(0)\tau}$
for small $\tau$. Consequently, the first two terms, $u_{0}\left(\frac{D}{2}-R(0)\tau\right)=2\lambda-\lambda R(0)\tau$,
constitute the Einstein-Hilbert term $\frac{R(\tau)}{16\pi G}$ when
$\tau$ is small at IR. The cosmological constant term $\lambda\nu$
provides a correction to Einstein gravity on the cosmic scale, while
those higher-order terms $O(R^{n}\tau^{n})$ offer corrections to
classical gravity at short distances.

The coupling constant $\lambda$, or equivalently the cosmic critical
density $\rho_{c}$, serves as the sole coupling constant for the
frame field in the quantum reference frame theory (\ref{eq:NLSM}),
and it is also the only characteristic energy scale of the frame field.
Therefore, the natural energy scale in quantum reference frame theory
is not the Planck scale, but rather the critical density energy scale
determined by the combination of the Hubble constant $H_{0}$ and
the Newton's constant $G$. The energy scales associated with the
cosmological constant and the critical density are the characteristic
energy scales of the quantum reference frame. This naturally explains
the cosmological constant problem, which is equivalent to explaining
why the characteristic scale of the universe differs so significantly
from the scale corresponding to the Newton's constant. The very low
energy scale of the gravitational system associated with the critical
density also implies that when the matter density becomes comparable
to this characteristic energy scale, gravitational behavior will significantly
deviate from the usual Einstein or Newtonian gravity \citep{Luo:2022ywl,Luo:2023eqf}.
For instance, noticeable modifications to Newtonian gravity emerge
at the low-density regions on the outskirts of spiral galaxies.

Shannon entropy/relative entropy, when serving as the effective action
for gravity, differs from the Einstein-Hilbert action in that the
functional of the Einstein-Hilbert action does not possess a gradient
flow. Instead, it generates a backwards flow that requires a continuous
input of new information during the flow process. Quantum corrections
cannot be absorbed into the coefficients of the original Einstein-Hilbert
action. If one attempts to eliminate these divergent quantum corrections
using conventional renormalization methods, it necessitates the continual
introduction of new terms and coefficients into the original action.
Moreover, if one proceeds to calculate quantum corrections for these
new terms and coefficients, even more new terms and coefficients must
be introduced to cancel out the previously calculated divergences.
This process appears to be endless. The absence of a gradient flow
is a significant reason why the so-called Einstein-Hilbert action
cannot be renormalized. However, the gradient flow of Shannon entropy
does exist and is a forward Ricci flow, which causes the gravitational
system to gradually average out short-distance scale information during
the flow process, with the entropy changing monotonically and non-decreasingly.
Eventually, the spacetime configuration flows to fixed points of extremal
entropy, namely, a finite number of GSRS (gradient shrinking Ricci
soliton) configurations. In this sense, the gravitational system is
renormalizable. The renormalizability of the gravitational system
essentially refers to the existence of long-flow-time solutions for
the Ricci flow, which flows toward a finite number of fixed-point
configurations, supplemented by controlled surgeries to overcome singularities
(the need for such surgeries arises because the Ricci flow only considers
corrections from second-moment fluctuations of the frame fields and
does not incorporate higher-order contributions, which become significant
near phase transitions and singularities).

\subsection{Schwarzschild Black Hole Entropy}

Another question is, given that the entropy of the gravitational system
and Lorentzian spacetime can be described using these monotonic entropies,
what is the relationship between these monotonic entropies and the
well-known thermodynamic entropies of gravitational systems, such
as the entropy of a Schwarzschild black hole?

Considering a stationary black hole with its mass distribution concentrated
at the coordinate origin, described by stress tensor $T_{00}=M\delta^{(3)}(\mathbf{x})$
and $T_{ij}=0$, $M$ the mass of the black hole. The Schwarzschild
black hole yields a scalar curvature $R(\mathbf{x})=-8\pi GT_{\mu}^{\mu}=8\pi GM\delta^{(3)}(\mathbf{x})$.
Substituting this into the Einstein equations, we find that the metric
at the origin in fact satisfies a temporal static and spacelike soliton
configuration (\ref{eq:GSRS})
\begin{equation}
R_{ij}(\mathbf{x})=8\pi GT_{ij}+\frac{1}{2}g_{ij}R(\mathbf{x})=\frac{1}{2}8\pi GM\delta^{(3)}(\mathbf{x})g_{ij}=\frac{1}{2\tau}g_{ij},\quad(i,j=1,2,3)
\end{equation}
in which the parameter $\tau$ behaves like Hawking's temperature
$\frac{1}{8\pi GM}$, and thus this soliton configuration corresponds
to a thermal equilibrium state with maximum entropy. To calculate
the Shannon entropy of a black hole, it is necessary to compute the
$u$ density on the Schwarzschild background either through its definition
(\ref{eq:4d u normalization}) or via the conjugate heat equation.
In the infrared limit, $u$ density is low frequency dominant, so
it can be simply viewed as a delta distribution centered around 3-momenttum
$|\mathbf{k}|=0$ with distributions on both sides, so it is almost
static and concentrated in the vicinity of the horizon $r_{H}$, approximating
a $\delta(r-r_{H})$ profile. So we approximately have the static
$u$ density distribution 
\begin{equation}
u_{\mathbf{k}}(r,\tau=0)\approx\delta(|\mathbf{k}|)\delta(r-r_{H})
\end{equation}
in the infrared limit, which gives a dominant contribution to the
Bekenstein-Hawking entropy, while those high-frequency modes (such
as those quasi-normal modes) all attenuate as the conjugate heat equation
evolves into the infrared limit. This delta function can be regarded
as being slightly broadened into a Gaussian profile at small $\tau$
by a evolution of the conjugate heat equation and hence symmetrically
distributed within a thin shell near the event horizon
\begin{equation}
u_{\mathbf{k}}(r,\tau)\approx\frac{1}{|\mathbf{k}|}\frac{1}{(4\pi\tau)^{1/2}}\exp\left[-\frac{(r-r_{H})^{2}}{4\tau}\right],\quad(\tau\rightarrow0)
\end{equation}
If the conjugate heat equation within (\ref{eq:4d coupled equs})
in this case is analogously regarded as a heat equation (temporally
static and spatically curved), then this solution similarly embodies
the maximum principle of the heat equation, namely, that the location
of the highest $u$ (temperature) appears at the boundary of the system
(the horizon seen from the rest observer at infinity) or occurs at
the initial moment when $\tau\rightarrow0$, or from another viewpoint,
the 4-volume element $\sqrt{|g|}$ archieves its minimum at the horizon.

Compared to the asymptotic flat background entropy $N_{*}$, the Shannon
entropy $N$ dominants the relative entropy $\tilde{N}$, so only
$N$ is considered in the following. Under the density profile given
above, first noting that $\log u_{\mathbf{k}}\approx-\frac{1}{2}\log\left(|\mathbf{k}|^{2}\tau\right)$,
we obtain the Shannon entropy for the k-modes
\begin{align}
N_{\mathbf{k}} & =-\int d^{3}Xu_{\mathbf{k}}\log u_{\mathbf{k}}\nonumber \\
 & =\delta(|\mathbf{k}|)\int_{r_{H}}^{\infty}4\pi r^{2}dr\frac{1}{(4\pi\tau)^{1/2}}\exp\left[-\frac{(r-r_{H})^{2}}{4\tau}\right]\frac{1}{2}\log(|\mathbf{k}|^{2}\tau)\nonumber \\
 & =\frac{1}{4}\delta(|\mathbf{k}|)A\log(|\mathbf{k}|^{2}\tau)
\end{align}
where $A=4\pi r_{H}^{2}$ is the area of the horizon. If we assume
that the radial momentum $k_{r}$ and the tangential momentum $k_{\bot}$
on the event horizon are uniformly isotropic, with $|\mathbf{k}|=|k_{r}|=|k_{\bot}|$,
then integrating the Shannon entropy over all k-modes with respect
to momentum yields the total entropy
\begin{align}
N & =\int d^{3}\mathbf{k}N_{\mathbf{k}}\nonumber \\
 & =\frac{1}{4}A\int\frac{d^{2}k_{\bot}}{(2\pi)^{2}}\log\left(|k_{\bot}|^{2}\tau\right)\int dk_{r}\delta(k_{r})\nonumber \\
 & =\frac{1}{4}A\int_{0}^{1/\epsilon}\frac{2\pi k_{\bot}dk_{\bot}}{(2\pi)^{2}}\log\left(|k_{\bot}|^{2}\tau\right)\nonumber \\
 & =\frac{1}{4}A\times\frac{1}{2\pi\tau}\left[-\frac{\tau}{2\epsilon^{2}}\left(1-\log\frac{\tau}{\epsilon^{2}}\right)\right]\nonumber \\
 & \approx-\frac{A}{16\pi\epsilon^{2}}
\end{align}
in which an ultraviolet cutoff $1/\epsilon$ is imposed on the integral
over the tangential momentum on the event horizon; otherwise, the
integral would diverge. In this way, we obtain a black hole entropy
that is proportional to the horizon area, weakly (logarithmically)
dependent on the flow parameter $\tau$ or temperature, and inversely
proportional to the square of the ultraviolet length cutoff $\epsilon$.
If we take the length cutoff to be the Planck length, i.e. $\epsilon^{2}=\frac{1}{4\pi}G$,
and define the thermodynamic entropy as the negative of the Shannon
entropy, we arrive at the Bekenstein-Hawking entropy.

\section{Conclusions}

In a covariant theory like gravity, where a global time does not exist,
conserved quantities such as energy in traditional dynamics are no
longer adequate as controls for the system. Instead, the Ricci flow
parameter $t$ (or $\tau$) serves as a semi-global parameter, and
its corresponding monotonic entropy functionals become crucial semi-global
control quantities for the system, in which the term ``semi-global''
means that the existence of the global parameter and global control
quantities (entropies) in the Lorentzian spacetime case can be persisted
for sufficient long flow-time, different from the stronger Riemannian
case which can be seen ``global'' as Perelman's monotonic functionals
had demonstrated. If the monotonic functionals for four-dimensional
Lorentzian spacetime proposed in this paper indeed exist, their implications
for gravitational systems, particularly quantum gravitational systems,
would be profoundly significant. They would play a crucial role in
controlling and constraining the quantum behavior of gravitational
systems. 

A key step in constructing monotonic functionals for four-dimensional
Lorentzian manifolds involves generalizing the positive-definite 4-volume
element and introducing a $u$ 4-density to the Lorentzian manifold
$(M,g,u)$, it leads to natural boundary conditions for the metric
$g$ and $u$ density. The $u$ density measures the local volume
comparison which is crucial because it encodes the important local
geometric information of a curved manifold, and so it measures gravity
such as the curvature emerge in the local volume comparison. The $u$
density defines a proper weighted inner product $\langle...\rangle=\int...\,u\sqrt{|g|}d^{4}X$
to the Lorentzian spacetime and choose a proper gauge in some operator
through the DeTurck trick (e.g. the Bakry-\'{E}mery curvature), which
is very important for the well-behaved properties of the spectrum
or eigenvalues of some operator in Lorentz signature. In certain sense,
introducing the new inner product defined by the $u$ density in Lorentzian
spacetime is analogous to defining an appropriate inner product via
second quantization through field variables, so that the negative
energy spectrum naturally present in Lorentzian signature is properly
handled. The $u$ density in the weighted Lorentzian spacetime eliminates
the unstable modes along non-compact timelike directions through the
gauge fixing introduced by the DeTurck trick, which is analogous to
the elimination of unstable negative modes of longitudinal photons
in the Faddeev--Popov gauge-fixed quantization.

Starting from the standard Shannon entropy given by the positive-definite
$u$ density, the first derivative of the Shannon entropy with respect
to the backwards flow parameter $\tau$ yields the generalized F-functional,
while the Legendre transform of the relative entropy gives the generalized
W-entropy functional in Lorentzian spacetime. These functionals for
four-dimensional Lorentzian spacetime bear a striking formal resemblance
to Perelman's functionals in the three-dimensional Riemannian case.
Another key step is to employ the DeTurck trick to choose an appropriate
gauge of the Ricci curvature, thereby ensuring the monotonicity of
the functionals over a sufficiently long flow-time. The monotonicity
of the generalized F- and W-functionals depends on the structure where
the derivative of them w.r.t. the flow parameter $t$ turns out to
be a perfect square of the Bakry-\'{E}mery curvature, which can be
non-negative for sufficient long flow-time by choosing an appropriate
gauge or DeTurck term to ensure the dominance of the real parts of
the complex eigenvalues of the Bakry-\'{E}mery curvature, and the
perfect square also means that it is independent of the sign of the
local curvature. In orther words, the real-part dominance achieved
through the DeTurck term, i.e. the Hessian, in gauge selection serves
as a guarantee for the monotonicity of the Ricci-DeTurck flow equation
for Lorentzian spacetime, and also in essential ensures the well-posedness
of the flow. The monotonicity of the F-functional in Lorentz spacetime
(in some sense) is equivalent to proving the positive energy theorem
at the Ricci flow or quantum level, which is also crucial for the
stability of the gravitational system. The gradient flows of these
functionals generate the Ricci flow for the four-dimensional Lorentzian
spacetime and the conjugate heat flow for the $u$ density under the
imposed normalization constraint (\ref{eq:4d u normalization}). Since
the Ricci flows of the timelike and spacelike modes of the metric,
as well as the conjugate heat flow of the $u$ density, are coupled
together, rather than isolated $u$ or $g_{\mu\nu}$ system, the entire
coupled system as the gradient flows of the monotonic functionals
is (semi-)globally controlled by these functionals. Although there
is no Sobolev inequality in Lorentz spacetime to guarantee the strict
boundedness of the initial functional in the mathematical rigor, but
at the level of physical rigor, the non-pathological nature of the
initial spacetime and the boundedness of the initial spacetime functional
can be ensured. Therefore, the facts that both the boundedness of
the initial functional and monotonicity of the functionals ensures
the Ricci flow and conjugate heat flow for the entire four-dimensional
Lorentzian spacetime are well-posed for sufficient long flow-time
at the physical rigorous level, ruling out that the timelike modes
of the metric and $u$ density experience unstably ``high-frequency
blow-up'' within a finite flow interval, otherwise, the functionals
would be divergent, which contradicts the monotonicity of the functionals
within a finite flow interval. 

In short, the Ricci-DeTurck flow equation in Lorentzian spacetime,
as a hyperbolic-parabolic system fixed by the DeTurck gauge-fixing
trick, has its long-flow-time behavior controlled by the dissipative
nature of the system or the real-part of the Bakry-\'{E}mery curvature.
The gauge freedom arising from the positive $u$ density, which comes
from the normalization (\ref{eq:4d u normalization}) by the positive
4-volume element and coupled to the spacetime and introduced in the
Bakry-\'{E}mery curvature, provides sufficient flexibility to ensure,
through an appropriate choice of gauge/coordinates, that the real-part
dominates over the imaginary part. This dominance allows the parabolic
dissipative behavior to prevail over the hyperbolic oscillatory behavior
in the long-flow-time scale, thereby ensuring the existence of the
monotonic functionals for the ($g_{\mu\nu}$ and $u$) coupled system
and the well-posedness of the Ricci-DeTurck flow equation in Lorentzian
spacetime, instead of the original Ricci flow without the DeTurck
term. Furthermore, just as the Faddeev-Popov gauge fixing confines
the configuration space of gauge fields to the Gribov physical subspace
of a specific gauge, the DeTurck trick's gauge fixing for the Bakry-\'{E}mery
curvature confines the Ricci-DeTurck flow to a physical configuration
subspace. In this physical subspace, the monotonicity of entropy ensures
that the local spacetime volume of the configuration does not collapse
within finite flow-time. The locally non-collapse theorem also guarantees
that, within a finite flow-time, either the Ricci-DeTurck flow cannot
reach the boundary of the physically configuration subspace, because
the boundary of the physical subspace is analogous to a Gribov boundary/horizon
where the Jacobian (i.e. Faddeev-Popov determinant) or the measure
collapse occurs, or as it approaches the boundary (developing singularity)
the flow is under control so that a Perelman-type surgery can also
be performed. Therefore, as long as an appropriate gauge is initially
selected through the DeTurck trick, the Ricci-DeTurck flow confined
in this entropy-monotonic physical configuration subspace seems always
be monotonic and hence well-posed for sufficiently long flow-time.
This mechanism also essentially ensures the well-posedness and monotonicity
of the Ricci--DeTurck flow of a Lorentzian spacetime within a proper
gauge fixing physical configuration subspace. Therefore, just as the
Faddeev-Popov gauge fixing eliminates the negative modes in quantum
gauge theories, within the DeTurck gauge-fixed physical configuration
subspace, the negative modes such as those exponentially blow-up unstable
modes in the Ricci-DeTurck flow are also eliminated, because the monotonicity
of the entropy functionals rule out their existence.

The Shannon entropy defined via the probability $u$ density of the
quantum frame fields, as one of these monotonic functionals, provides
the dominant contribution to the quantum gravitational partition function.
The Shannon entropy of the quantum frame field describes anomalies
in general quantum coordinate transformations, which asymptotically
recovers the classical Einstein-Hilbert action in the infrared limit.
The anomaly cancellation term yields the correct cosmological constant.
The Shannon entropy corresponding to the $u$ density in the Schwarzschild
spacetime background gives the area law of the Bekenstein-Hawking
entropy. The $u$ density, as the probability density giving rise
to the spacetime Shannon entropy, describes the ensemble density of
the quantum frame field as the microscopic degrees of freedom of spacetime.
This reflects that spacetime is a general equivalent, which is an
abstraction of the quantum frame fields, that simplifies the studying
of the relative motion between matters, according to the quantum equivalence
principle.
\begin{acknowledgments}
This work was supported in part by the National Science Foundation
of China (NSFC) under Grant No.11205149, and the Scientific Research
Foundation of Jiangsu University for Young Scholars under Grant No.15JDG153.
\end{acknowledgments}

\bibliographystyle{unsrt}

\end{document}